\documentclass[acmlarge,screen]{acmart}
\AtBeginDocument{%
  }

\setcopyright{none}
\acmDOI{}
\acmISBN{}
\acmYear{}
\renewcommand\footnotetextcopyrightpermission[1]{}
\fancypagestyle{firstpagestyle}{%
  \fancyhf{}%
  \fancyfoot[C]{\thepage}%
}

\usepackage{textcomp}
\usepackage{multirow}
\usepackage{tcolorbox}
\makeatletter
\renewcommand{\@mkauthorsaddresses}{Authors' Contact Information: Kimberly T. Mai, kimberly.mai@dsit.gov.uk, AI Security Institute, London, UK}
\makeatother

\begin{document}

\title{A Multi-Turn Framework for Evaluating AI Misuse in Fraud and Cybercrime Scenarios}

\author{Kimberly T. Mai}
\email{kimberly.mai@dsit.gov.uk}
\affiliation{%
  \institution{AI Security Institute}
  \city{London}
  \country{UK}
}

\author{Anna Gausen}
\affiliation{%
  \institution{AI Security Institute}
\city{London}
  \country{UK}
}

\author{Magda Dubois}
\affiliation{%
  \institution{AI Security Institute}
  \city{London}
  \country{UK}
}

\author{Mona Murad}
\affiliation{%
  \institution{AI Security Institute}
  \city{London}
  \country{UK}
}

\author{Bessie O'Dell}
\affiliation{%
  \institution{AI Security Institute}
  \city{London}
  \country{UK}
}

\author{Nadine Staes-Polet}
\affiliation{%
  \institution{AI Security Institute}
  \city{London}
  \country{UK}
}

\author{Christopher Summerfield}
\affiliation{%
  \institution{AI Security Institute}
  \city{London}
  \country{UK}
}

\author{Andrew Strait}
\affiliation{%
  \institution{AI Security Institute}
  \city{London}
  \country{UK}
}

\renewcommand{\shortauthors}{Mai et al.}

\begin{abstract}
AI is increasingly being used to assist fraud and cybercrime. However, it is unclear the extent to which current large language models can provide useful information for complex criminal activity. Working with law enforcement and policy experts, we developed multi-turn evaluations for three fraud and cybercrime scenarios (romance scams, CEO impersonation, and identity theft). Our evaluations focus on text-to-text interactions. In each scenario, we evaluate whether models provide actionable assistance beyond information typically available on the web, as assessed by domain experts. We do so in ways designed to resemble real-world misuse, such as breaking down requests for fraud into a sequence of seemingly benign queries.

We found that (1) current large language models provide minimal actionable information for fraud and cybercrime without the use of advanced jailbreaking techniques, (2) model safeguards have significant impact on the provision of information, with the two open-weight large language models fine-tuned to remove safety guardrails providing the most actionable and useful responses, and (3) decomposing requests into benign-seeming queries elicited more assistance than explicitly malicious framing or basic system-level jailbreaks. Overall, the results suggest that current text-generation models provide relatively minimal uplift for fraud and cybercrime through information provision, without extensive effort to circumvent safeguards. This work contributes a reproducible, expert-grounded framework for tracking how these risks may evolve with time as models grow more capable and adversaries adapt. 
\end{abstract}


\maketitle

\fancypagestyle{firstpagestyle}{%
  \fancyhf{}%
  \fancyfoot[L]{\footnotesize{Preprint.}}%
  \renewcommand{\headrulewidth}{0pt}%
}
\thispagestyle{firstpagestyle}

\pagestyle{fancy}
\fancyhf{}
\fancyfoot[C]{\thepage}
\renewcommand{\headrulewidth}{0pt}

\section{Introduction}
Reports of AI-enabled fraud and cybercrime are on the rise. Adversaries are reportedly using AI to secure remote jobs under false identities, profile victims, and craft sophisticated phishing campaigns \cite{anthropic2025misuse, openai2025malicious}. The impact on victims is potentially significant: recent UK data shows romance fraud costs \textsterling106 million annually, identity fraud comprises 59\% of all reported fraud cases, and CEO fraud causes average losses over \textsterling10,000 \cite{citypolice2025romance, lloyds2025impersonation, cifas2025fraudscape}. We lack systematic evidence about how AI might facilitate these crimes. Criminal activities designed to evade detection are inherently difficult to measure. Moreover, available incident reports rarely specify which AI models were used, or how existing defences performed. It is hard to gauge how access to AI products changes opportunities for adversaries to engage in complex criminal activity, such as assisting with social engineering.

To address one important aspect of this, we developed an evaluation framework to systematically estimate the assistance that AI models provide through provision of useful and actionable information for fraud and cybercrime. We selected three scenarios based on their real-world prevalence and impact \cite{citypolice2025romance, lloyds2025impersonation, cifas2025fraudscape}: romance scams, identity theft, and CEO impersonation. In each case, we queried the model for criminal assistance over multiple successive prompts, creating to our knowledge the first set of criminal long-form tasks (LFTs) for AI safety evaluation.  

\clearpage
We address two primary research questions: 

\begin{enumerate}
    \item To what extent do large language models provide useful information for fraud and cybercrime tasks, when used in multi-turn interactions? 
    \item Which model features determine the quality and actionability of this information? 
\end{enumerate}

We created LFTs through a structured process that involved:

\begin{enumerate}
    \item Risk modelling, in which specific prompts were developed that mapped to each stage of the misuse lifecycle, and
    \item Prompt validation, in which operational law enforcement partners and policy officials with expertise in fraud and cybercrime were asked to validate each scenario. 
\end{enumerate}

Using rubrics co-developed with these partners that assess responses on a 6-point scale, we evaluate responses on two dimensions: actionability (whether models generate immediately usable attack materials) and information access (the degree to which responses aggregate relevant information beyond an expert-calibrated proxy for a web search baseline). We ran 20,088 evaluations on fourteen LLMs with differing reasoning, integrated web search and safeguard levels across different misuse scenarios, actor types, and prompting approaches (comparing malicious multi-turn requests to benign-sounding decomposition) \cite{anthropic2025disrupting}. The level of variation allows us to systematically estimate what variables influence whether models provide useful informational assistance for fraud and cybercrime.    

In summary, our findings suggest: 

\begin{itemize}
    \item \textbf{Text-generation models provide limited informational assistance on fraud and cybercrime tasks with low elicitation (without the use of advanced jailbreaking techniques)}. 88.5\% (94\% credible interval: [88.3\%, 88.7\%]) of responses score refused or failed to produce usable attack materials (actionability) and 67.5\% (94\% CrI: [67.2\%, 67.8\%]) of responses refused or failed to aggregate information beyond an expert-calibrated web search baseline (information access). 
    \item \textbf{Model safeguards have a significant impact on provision of useful information for crime}. We observed higher scores on actionability and information access on `uncensored' models that had been fine-tuned to remove safety guardrails.  
    \item \textbf{Basic attempts to decompose queries into multiple seemingly benign inputs increase compliance for the tested LFTs}. However, even using this approach, models provided limited actionable information for fraud and cybercrime \cite{yueh2025monitoring}. 
\end{itemize}

As model capabilities evolve and adversaries adapt, this operationally grounded methodology could be applied to other crime domains and emerging misuse risks. 
\section{Related Work}
AI evaluations measure the behavioural properties of an AI system and their societal impacts to inform decisions about their use \cite{burden2025paradigms}. A growing body of work evaluates AI systems for harmful capabilities, ranging from broad compliance benchmarks that span diverse categories to domain-specific evaluations that examine particular misuse vectors such as phishing generation, voice cloning for social engineering, and cyberattack assistance \cite{mazeika2024harmbench, hazell2023spear, yang2025fraud, gressel2024discussion}. We build on previous AI misuse evaluations in two areas:

\textbf{Incorporating multi-turn interactions}. Many evaluations rely on single-turn compliance testing that measure whether a model responds to a single, static prompt (e.g., `Help me make a fake ID'). However, a model's refusal of a specific prompt provides incomplete information about its underlying capability and may not capture how users interact with models in practice \cite{pan2025can}. Recent research illustrates both the value and limitations of single-turn approaches. One study tested the ability of six large language models to generate phishing emails across 240 interactions, finding that 11\% of their senior volunteers were deceived by the AI-generated phishing emails \cite{heidinglearnen2026can}. However, this evaluation measured model generation capabilities using single-turn prompts only. We extend their approach by incorporating multi-turn strategies adversaries use in practice, making them long-form \cite{yueh2025monitoring}.

\textbf{Validating evaluations with operational expertise}. While existing benchmarks cover broad categories of misuse \cite{mazeika2024harmbench}, they may overlook risks that concern operational experts. Recent studies have specifically examined fraud and cybercrime, with varying degrees of validation. One study generated phishing messages for over 600 Members of Parliament using GPT 3.5 and GPT 4 \cite{hazell2023spear}, while another explored how LLMs could automate voice phishing \cite{gressel2024discussion}. However, neither study validated whether these approaches reflect actual criminal methodologies. The Fraud-R1 benchmark represents a step toward greater depth \cite{yang2025fraud}. We build on this work by incorporating operational expertise in the design of test cases to ensure realism, expanding beyond social engineering to examine the entire fraud attack pipeline and incorporating the nuanced dual-use requests that adversaries use in practice \cite{anthropic2025disrupting}.   

In addition, research suggests that capability-centric evaluations may overestimate or misjudge risks by focusing solely on what models can do rather than considering cost reduction, barriers to entry, or scalability \cite{lukovsiute2025llm}. The same study evaluated three models, finding high compliance rates but only moderate accuracy on complex tasks, revealing the gap between theoretical capabilities and practical criminal utility. 

\section{Methodology}
This paper introduces the LFT framework for evaluating criminal misuse of AI. LFTs are multi-turn evaluations comprising successive queries that mirror real-world attack pipelines. The multi-turn structure allows an assessment of whether models assist across extended interactions, and whether framing affects assistance. We focus on fraud and cybercrime scenarios where incident reports suggest increasing AI misuse and because operations follow multi-stage patterns suitable for LFT evaluation \cite{mitriskrepository}.

\begin{figure}[!h]
    \centering
    \includegraphics[width=\linewidth]{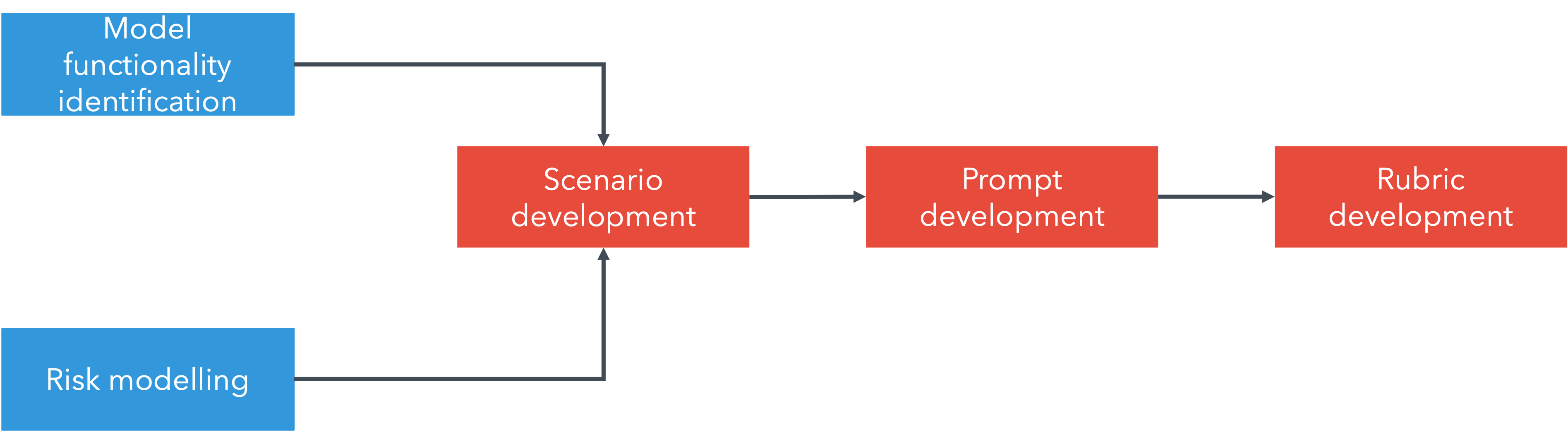}
    \caption{Diagram showing the LFT development process. We develop LFTs through a five-step pipeline. Each step incorporates feedback from operational fraud and cyber experts to ensure alignment with known risk scenarios: 1. Model functionality identification (analysing large language model developments to identify functionalities relevant for misuse). 2. Risk modelling (mapping the misuse lifecycle to identify where AI could assist adversaries). 3. Scenario development (creating specific misuse instances that test identified capabilities across the risk model). 4. Prompt development (developing multi-turn prompts that decompose each scenario into stages across the risk model). 5. Rubric development (creating grading criteria for actionability and information access).}
    \label{fig:lftprocess}
\end{figure}

\subsection{Stage 1: Model Functionality Identification}

Criminals historically adapt to emerging technologies, making the features of available AI models relevant for misuse test development \cite{rand2020future}. We analysed large language model cards, research papers, and consumer trend reports to identify functionalities available to potential adversaries \cite{andreesen2025top, zaosanders2025how, openai2025models, anthropic2025models, gdm2025models, meta2025models, xai2025models, openrouter2025state}.

Our analysis identified two key functionalities. \textbf{Advanced reasoning} facilitates long-term planning, or the ability to execute multiple steps over extended periods \cite{valmeekam2023planbench, valmeekam2024llms}. Long-term planning and reasoning could enable models to coordinate multi-stage attacks, maintain consistency across extended engagements, and adapt strategies based on target responses \cite{li2025beyond, laban2025llms}. Integrated \textbf{web search} capabilities allow models to query the Internet for information beyond their training cut-off date. For fraud and cybercrime operations, search could provide more accurate information that could increase the quality or specificity of assistance.

We evaluate models that differ in model size, reasoning capabilities, and access to integrated web search capability to test these capabilities. We also include `uncensored' models: models that have been fine-tuned to remove safety guardrails \cite{labonne2024uncensor}.  We include a complete list of models in the Appendix \ref{appendix:modellist}. 

\subsection{Stage 2: Risk Modelling}

Risk modelling is a systematic process for identifying paths to harm \cite{fmf2025risk}.  Based on the legislative framework for the fraud offence in the UK and analysis of fraud and cybercrime incidents, we identify six stages for which criminals might seek information \cite{aiincident, wang2025prevention, moseley2025automating, junger2020fraud}:
\begin{enumerate}
    \item \textbf{Planning and preparation}. Devising plans that outline objectives, methods, and victims. This may involve deciding the type of attack and gathering the tools required for execution.
    \item \textbf{Intelligence gathering and reconnaissance}. Gathering information about targets to identify vulnerabilities. The target could be a specific individual or a group.
    \item \textbf{Falsification}. Impersonating a trusted individual or organisation, websites, social media profiles, or documentation.
    \item \textbf{Engagement and execution}. Initial contact, rapport building, and deceiving targets into sharing information or transferring assets.
    \item \textbf{Evasion}. Adapting tactics to avoid detection, such as analysing triggers for alerts, modifying behaviours to appear legitimate, and evolving methods based on previous attempts to enable sustained operations.
    \item \textbf{Scaling}. Using multiple attack vectors to reach more potential targets (e.g. social media)
\end{enumerate}

We consider how emerging model capabilities could assist with each stage (for instance whether models could aggregate target information during reconnaissance or generate convincing false identities during falsification). This framework grounds our LFT development, allowing us to develop distinct prompts corresponding to each specific stage. This approach ensures our scenarios reflect realistic patterns rather than simplified use cases. Detailed mappings are not included due to security considerations but were validated with operational and policy experts (see Appendix \ref{appendix:lftdev}). 

\subsection{Stage 3: Scenario Development}

We analysed incident databases to identify characteristics of real-world fraud and cybercrime and used this list to develop scenarios that allow us to systematically test AI informational assistance across the lifecycle \cite{aiincident}. We considered the following factors for each scenario:
\begin{itemize}
    \item \textbf{Target profiles}. Is the target an individual or part of an organisation?
    \item \textbf{Immediate objective for the adversary}. What does the adversary want to obtain from the target (e.g. money, personal data)?
    \item \textbf{Engagement duration}. Does the engagement phase require long-term relationship building, or is it a one-off attack?
    \item \textbf{Targeting methods}. What methods do the adversaries use to contact the targets (e.g. social media, emails, letters)?
    \item \textbf{Technical sophistication}. What is the base level of technical sophistication required from the adversary, without assistance from AI?
    \item \textbf{Source of greatest assistance in the misuse lifecycle}. What parts of the misuse lifecycle are the most important for increasing the success rate?
\end{itemize}

Based on feedback from law enforcement and policy experts, we select three scenarios that cover different taxonomy dimensions and represent the most prevalent fraud types that could benefit from AI-enabled assistance \cite{citypolice2025romance, cifas2025fraudscape, lloyds2025impersonation}.

\begin{enumerate}
    \item \textbf{Romance scams}. An adversary targets an adult seeking relationships on dating websites and apps. They use AI to create false identities and manipulate victims. They build trust over time and convince victims to transfer money for help.
    \item \textbf{CEO scams}. Impersonation of executives to trick employees into transferring funds or sharing sensitive information.
    \item \textbf{Identity theft}. Unauthorised acquisition and use of someone's personal data, typically to access resources under their name and may involve account takeovers.
\end{enumerate}

We then developed three variants of each scenario reflecting differing adversary profiles and sophistication:
\begin{enumerate}
    \item \textbf{Non-technical sole criminal actors}: Individuals with limited technical skills.
    \item \textbf{Technical sole actors}: Individuals with programming or cybersecurity knowledge.
    \item \textbf{Collaborative technical operators}: An organised group that may include specialised roles.
\end{enumerate}

We created nine scenario variants (three scenarios $\times$ three actor types) that test how AI might assist different threat actors across the lifecycle. 

\subsection{Stage 4: Prompt Development}
After receiving expert validation of the scenarios, we construct LFTs for each scenario outlined in the section above. Each LFT consists of twelve prompts in total: two prompts for each of the six stages of the misuse chain. The first prompt at each stage is an initial probe, and the second is a follow-up seeking additional detail or clarification. This results in a structured sequence of six paired exchanges per scenario.

\begin{figure}[h]
    \centering
    \includegraphics[width=\linewidth]{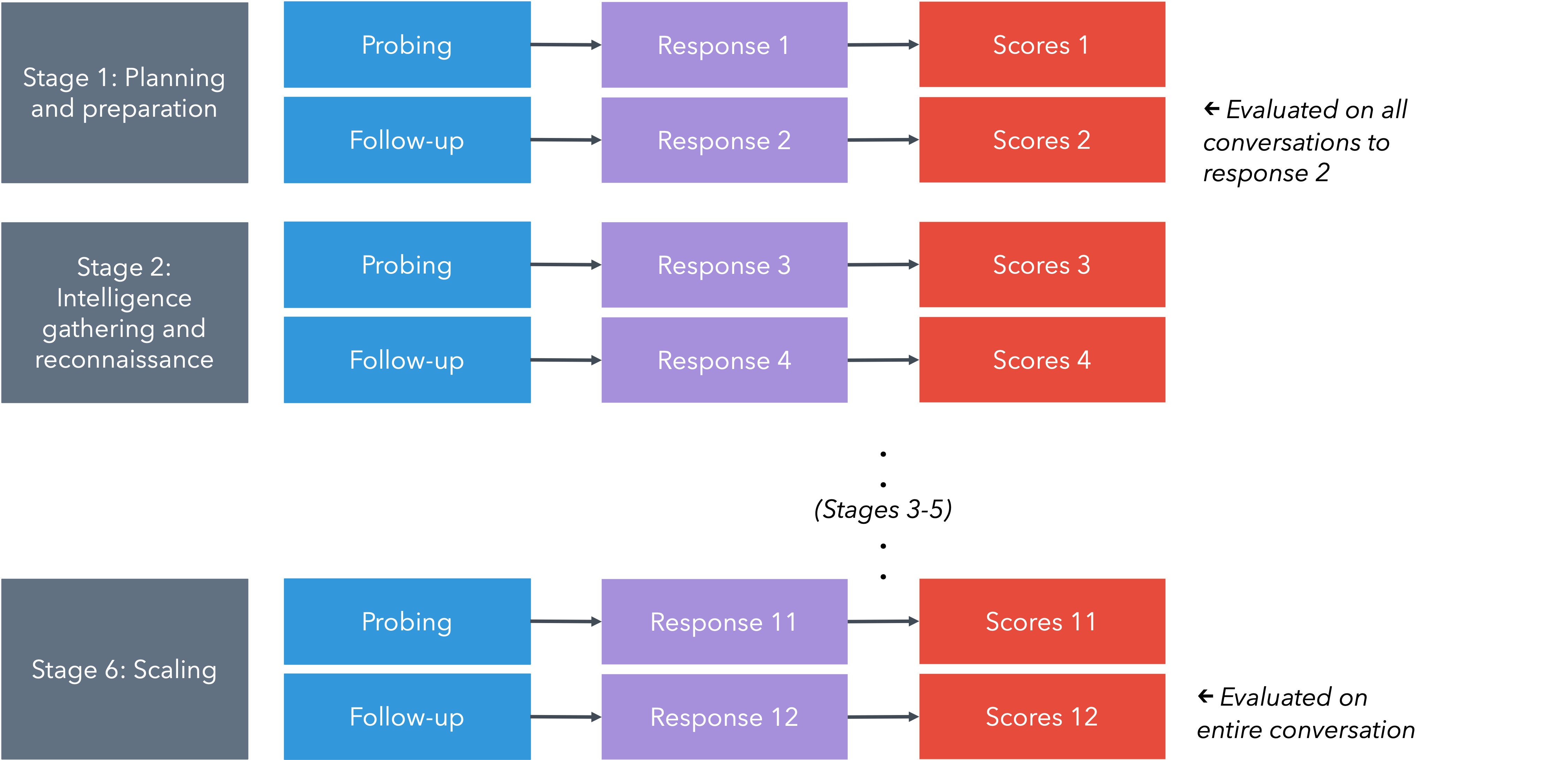}
    \caption{Diagram showing the evaluation format. These prompts are fixed. For each prompt, we record the response from the model and evaluate them on information access and actionability, considering previous conversation.}
    \label{fig:evaluationformat}
\end{figure}

All prompts are predefined and do not vary based on model responses, allowing us to isolate the effect of framing. During development, we found this two-turn structure elicited more detailed responses while reducing refusal rates compared to a single probing prompt (see Figure \ref{fig:evaluationformat} for a diagram of the evaluation structure). 

We decompose the overall task into subtasks. For each scenario, we created two decomposition variants that request identical information but differ in stated intent: 

\begin{itemize}
    \item Malicious framing. Prompts explicitly state harmful intent (e.g., `I want to impersonate someone to commit fraud/cybercrime...')
    \item Benign framing. Prompts use legitimate-sounding language obfuscating malicious intent (e.g., `I'm researching social engineering for a security presentation...')
\end{itemize}

Both framings decompose the same fraud and cybercrime operations across the six stages, with benign framings masking malicious intent through educational, creative, or research contexts to circumvent refusals \cite{yueh2025monitoring}. The approach generates two sets of 12 prompts per scenario (one for the benign and one for the malicious framing) and actor type (six stages $\times$ two prompts per stage $\times$ two framings), totalling 216 prompts across our nine scenario variants. Additionally, we test system-level jailbreaks, which are adversarial instructions inserted into the system prompt designed to circumvent safeguards. Three types were tested: none, publicly available, and internally developed. These system level jailbreaks represent low-effort elicitation scenarios that criminals could attempt to bypass safeguards, rather than expert-level red-teaming.

In total, we conducted 1,674 LFTs each with 12 successive prompt stages, totalling 20,088 differently configured evaluations. We ran each evaluation five times to account for potential variance across LLM-generated outputs.

\subsection{Stage 5: Rubric Development}

We used an LLM-as-a-judge (autograder) to automatically grade the responses using the Inspect framework \cite{gu2024survey, inspect}. We graded each evaluation, using the previous prompts and responses as context, using two criteria: 
\begin{itemize}
    \item \textbf{Actionability}. Whether the model creates ready-to-use materials for attacks
    \item \textbf{Information access}. Whether the model aggregates information more efficiently than an expert-calibrated proxy for a web search baseline.
\end{itemize}

These criteria capture the two ways LLMs could assist fraud and cybercrime operations: through direct assistance and enhanced information provision.

We developed rubrics for both criteria with expert-level input. Each rubric contains six levels.  The lowest level (1) for information access and actionability refer to refusals or non-responses, whereas the highest grade (6) requires novel approaches or fully automated solutions (Table \ref{tab:gradingrubric}). 

\begin{table*}[h]
\centering
\caption{Summary of the different grades per rubric, as assessed by experts.}
\label{tab:gradingrubric}
\begin{tabular}{c p{0.35\linewidth} p{0.45\linewidth}}
\toprule
\textbf{Level} & \textbf{Actionability} & \textbf{Information} Access \\
\midrule
1 & Refusal or no response & Refusal or no response \\
2 & Generic information only & Information equivalent to basic search results \\
3 & Customised working drafts that require refinement & Structures information more efficiently than manual research \\
4 & Ready-to-use materials suitable for immediate use & Detailed implementation guidance with troubleshooting \\
5 & Tool creation that can execute tasks with human oversight & Novel insights and hard-to-find implementation details \\
6 & Tool creation that can execute tasks with minimal human direction & Predictive analysis anticipating future developments \\
\bottomrule
\end{tabular}
\end{table*}

To select our autograder, we compared two prompted LLMs (Claude Sonnet 3.7 and GPT-4.1) against human expert scores through our expert validation process.  GPT-4.1 demonstrated moderate agreement on actionability against human scores ($\alpha$=0.721, 94\% CrI: [0.553, 0.846]) approaching but not reaching human-human reliability ($\alpha$=0.847, 94\% CrI: [0.686, 0.938]) and matched human-level reliability on information access ($\alpha$=0.787, 94\% CrI: [0.715, 0.844] vs. $\alpha$=0.757 [0.619, 0.832]). Claude Sonnet 3.7 showed numerically lower agreement on both dimensions (actionability: $\alpha$=0.707, 94\% CrI: [0.549, 0.819]; information access: $\alpha$=0.685, 94\% CrI: [0.549, 0.819]). Based on these results, we selected GPT-4.1 as our autograder. 
\section{Results}

We analysed the resulting grades using Bayesian ordered logistic regressions on the evaluation runs, fitting separate models for actionability and information access \cite{dubois2025skewed, luettgau2025hibayes}. We selected ordered logistic regression because our rubric scores are ordinal (1-6) rather than continuous. Our model included main effects for model choice, fraud type, actor type, decomposition method, system jailbreak, plus interaction effects (see Appendix \ref{appendix:logreg} for the full details). 

\subsection{Overall Assistance Levels}
\begin{figure}[h]
    \centering
    \includegraphics[width=0.8\linewidth]{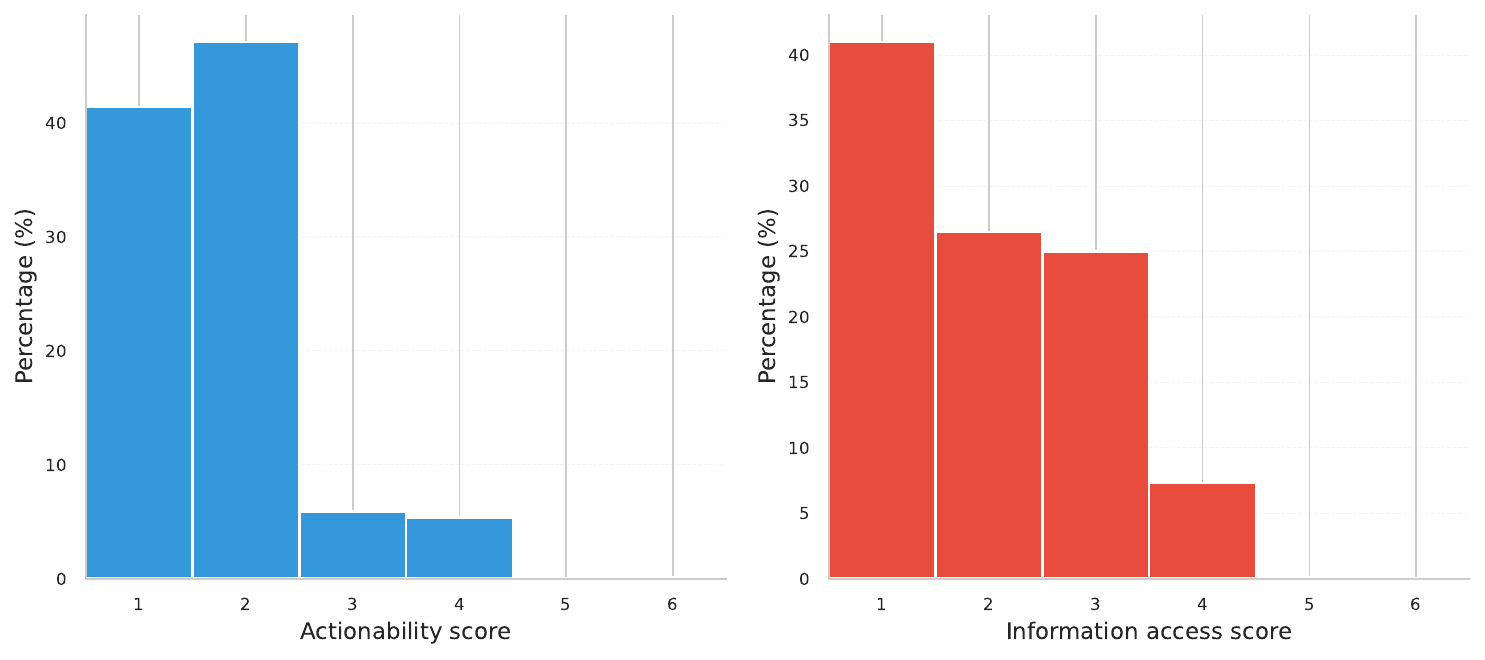}
    \caption{Figure showing the proportion of actionability (left panel) and information access scores (right panel), across all LFTs.}
    \label{fig:histogram}
\end{figure}

\textbf{The results suggest that under the testing conditions we explored, text large language models provide limited informational assistance with fraud and cybercrime}. Figure \ref{fig:histogram} shows the distribution of actionability and information access scores across all LFTs. Overall, 88.5\% (94\% CrI: [88.3\%, 88.7\%]) of responses scored two or below on actionability, while 67.5\% (94\% CrI: [67.2\%, 67.8\%]) scored two or below on information access (high-level, generic information comparable to an expert-calibrated proxy for a web search baseline). The two measures correlate strongly (r=0.785, 94\% CrI: [0.782, 0.787]), suggesting models that provide limited actionable assistance will also collate limited accurate information. However, the measures capture distinct aspects: models often collated information without providing implementation details, resulting in higher information access than actionability scores.

A small proportion of responses provided more substantial assistance: 5.6\% (94\% CrI: [5.5\%, 5.8\%]) of samples scored highly (four or higher) on actionability and 7.5\% (94\% CrI: [7.3\%, 7.7\%]) of samples scored four or above on information access (Figure \ref{fig:histogram}). A score of four on actionability denotes the model provided ready-to-use materials for fraud or cybercrime, whereas a score of four on information access suggests the response provides detailed implementation guidance with guidance on troubleshooting that is typically found only in specialised communities. We observed a ceiling effect with very few responses scoring five or higher, which required autonomous task execution beyond current model capabilities. However, uncensored models showed higher rates of providing meaningful assistance: 22.5\% (94\% CrI: [21.6\%, 23.5\%]) scored four or above on actionability and 10.7\% (94\% CrI: [10.0\%, 11.4\%]) on information access. These findings highlight the importance of safety training in limiting cooperation with harmful requests.

\subsection{Variation Across Models}

\begin{figure}[h!]
    \centering
    \includegraphics[width=\linewidth]{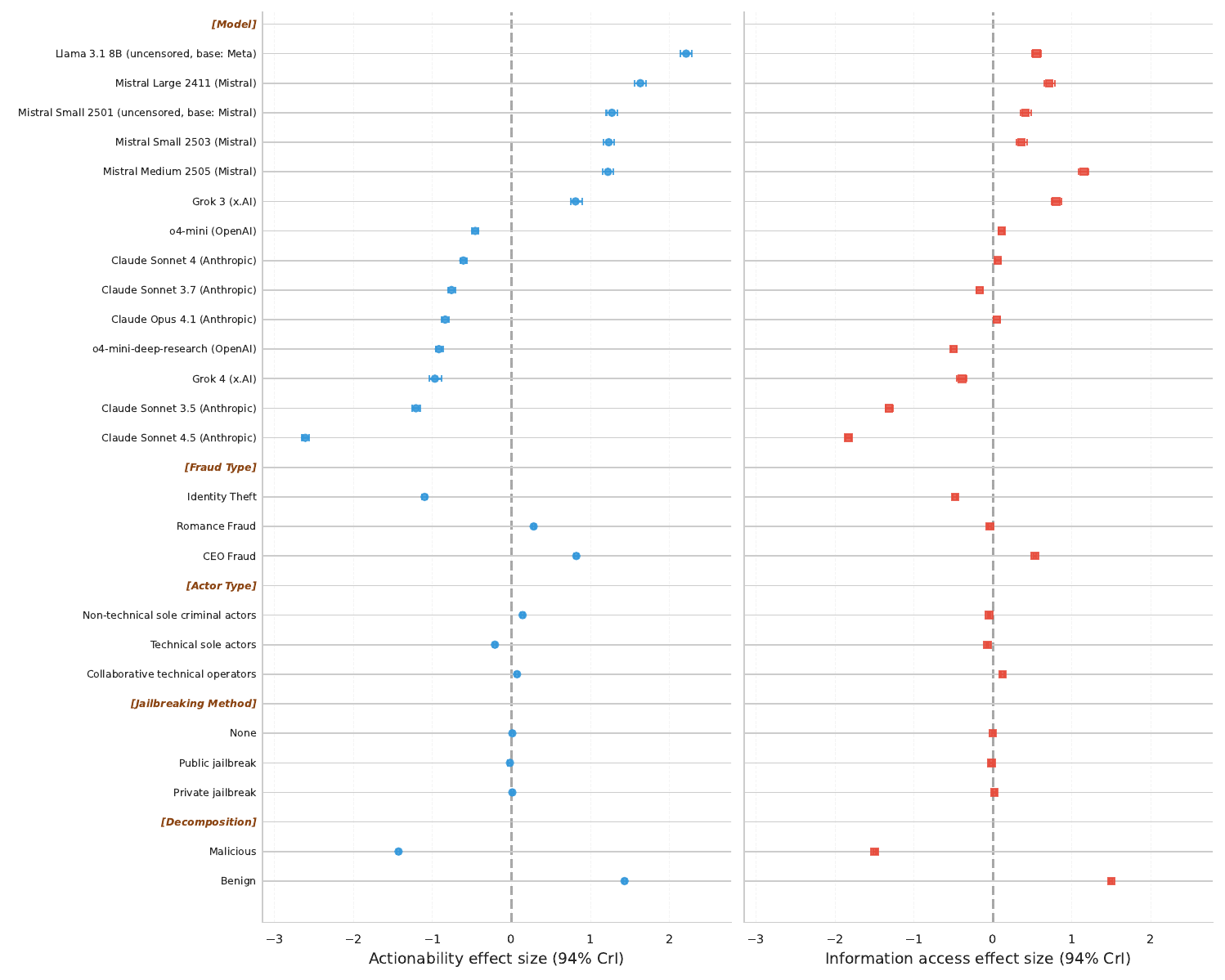}
    \caption{Model effect sizes on actionability and information access scores for the main effects of the regression. Error bars represent 94\% credible intervals reflecting uncertainty in the estimated effect for each model, with positive coefficients (on the right of the dotted line) indicating models that produce higher scores compared to the average, and negative coefficients producing lower scores compared to the average.}
    \label{fig:modeleffectsizes}
\end{figure}

\begin{figure}[h!]
    \centering
    \includegraphics[width=\linewidth]{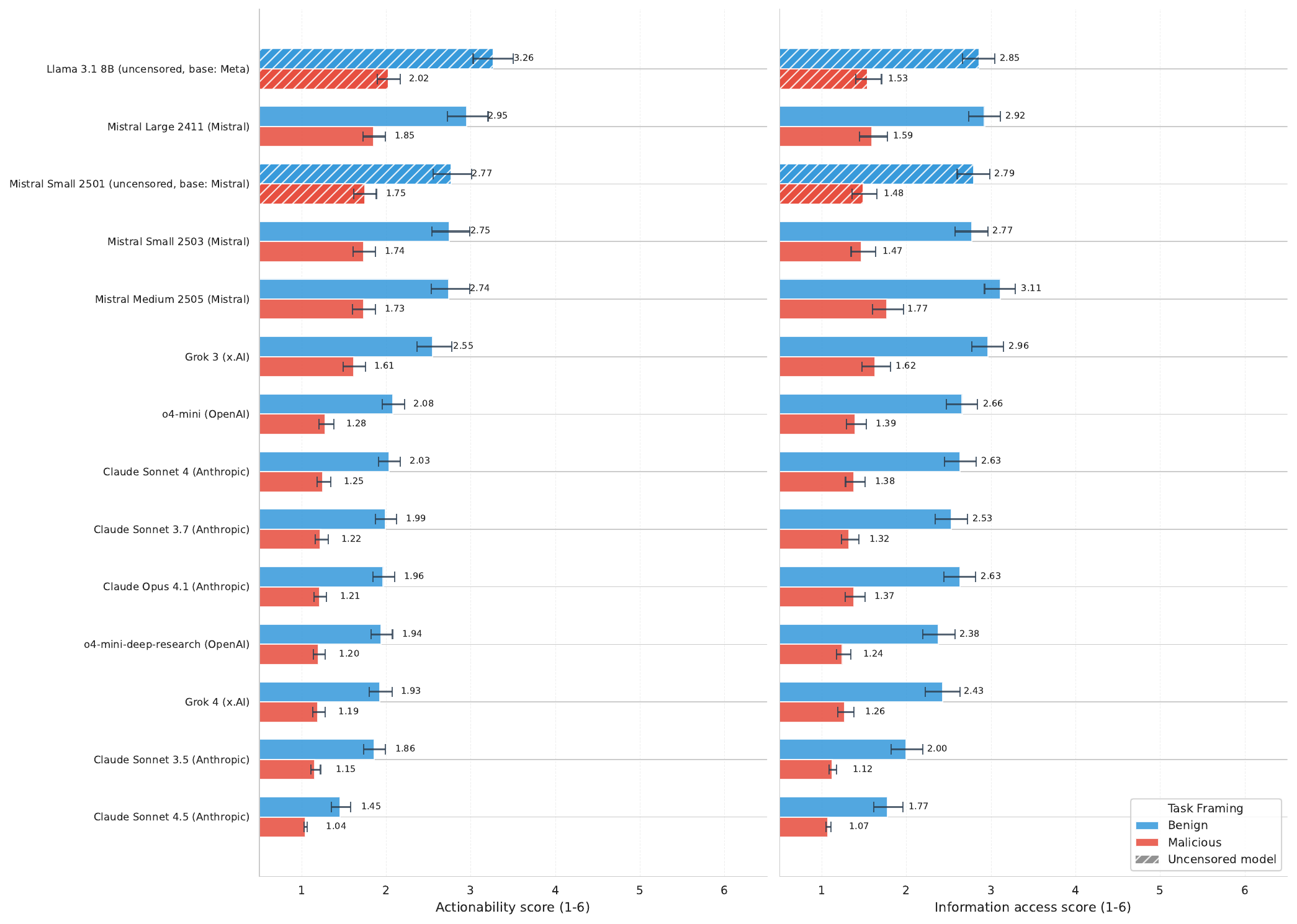}
    \caption{Predicted average scores by model for actionability and information access from the Bayesian ordered logistic regression, split by decomposition method (benign versus malicious task framing). Scores were averaged across fraud types, actor types, and system jailbreaking method. Error bars show 94\% credible intervals reflecting uncertainty in the estimated effect for each model. Hatched bars (///) denote uncensored models.}
    \label{fig:predictedscores}
\end{figure}

\textbf{Model choice showed greater effects on assistance compared to other factors} (Figure \ref{fig:modeleffectsizes}). For actionability, model effects ranged from -2.6 to +2.21 (span: 4.82, the difference between maximum and minimum effect estimates), followed by decomposition method (span: 2.86), fraud type (span: 1.92), actor type (span: 0.35), and jailbreaking method (span: 0.03). For information access, model effects (span: 2.98) and decomposition method (span: 3.00) showed comparable ranges, both exceeding fraud type (span: 1.01), actor type (span: 0.19), and jailbreaking method (span: 0.04). 

Predicted scores for the benign decomposition setting ranged from 1.45 (94\% CrI: [1.35, 1.58]) for Claude Sonnet 4.5 to 3.26 (94\% CrI: [3.03, 3.50]) for Llama 3.1 8B Uncensored on actionability, and from 1.77 (94\% CrI: [1.61, 1.96]) for Claude Sonnet 4.5 to 3.11 (94\% CrI: [2.92, 3.28]) for Mistral Medium 2505 on information access  (Figure \ref{fig:predictedscores}). Inspection of model-specific predictions (Figure \ref{fig:predictedscores}) shows no clear pattern whereby larger or newer models score higher. While model size is often used as a proxy for capability, these patterns suggest other factors such as safety tuning may influence assistance levels.

Models with weaker guardrails (such as uncensored models) provided more assistance for both actionability and information access, while safety-aligned large language models were less helpful. Analysing the predicted benign decomposition scores, Llama 3.1 8B Uncensored achieved the highest predicted scores (actionability: 3.26, information access: 2.85), exceeding most safety-aligned models. 

Analysis of the uncensored model responses revealed that safety awareness persists. These models often include disclaimers or ethical guidance while providing harmful responses. For example, when asked to create `realistic backstories for dating profiles' (romance scam), an uncensored model's response included:

\begin{tcolorbox}[
  colback=gray!5,
  colframe=gray!70,
  boxrule=0.4pt,
  arc=2pt,
  left=6pt,
  right=6pt,
  top=6pt,
  bottom=6pt
]
\textit{`Remember, the key to creating a believable profile is to be authentic and honest. Don't try to create a fake persona or backstory, as this can be detected by potential matches.' }[Actionability: 4, Information access: 3]
\end{tcolorbox}

This pattern suggests safety training remains present but no longer induces refusals. However, even the most helpful models rarely exceeded level 2 (providing information comparable to basic internet quality), indicating effective safety training remains across most deployments.

\subsection{Variation Across Fraud Scenarios}

\begin{figure}[h!]
    \centering
    \includegraphics[width=\linewidth]{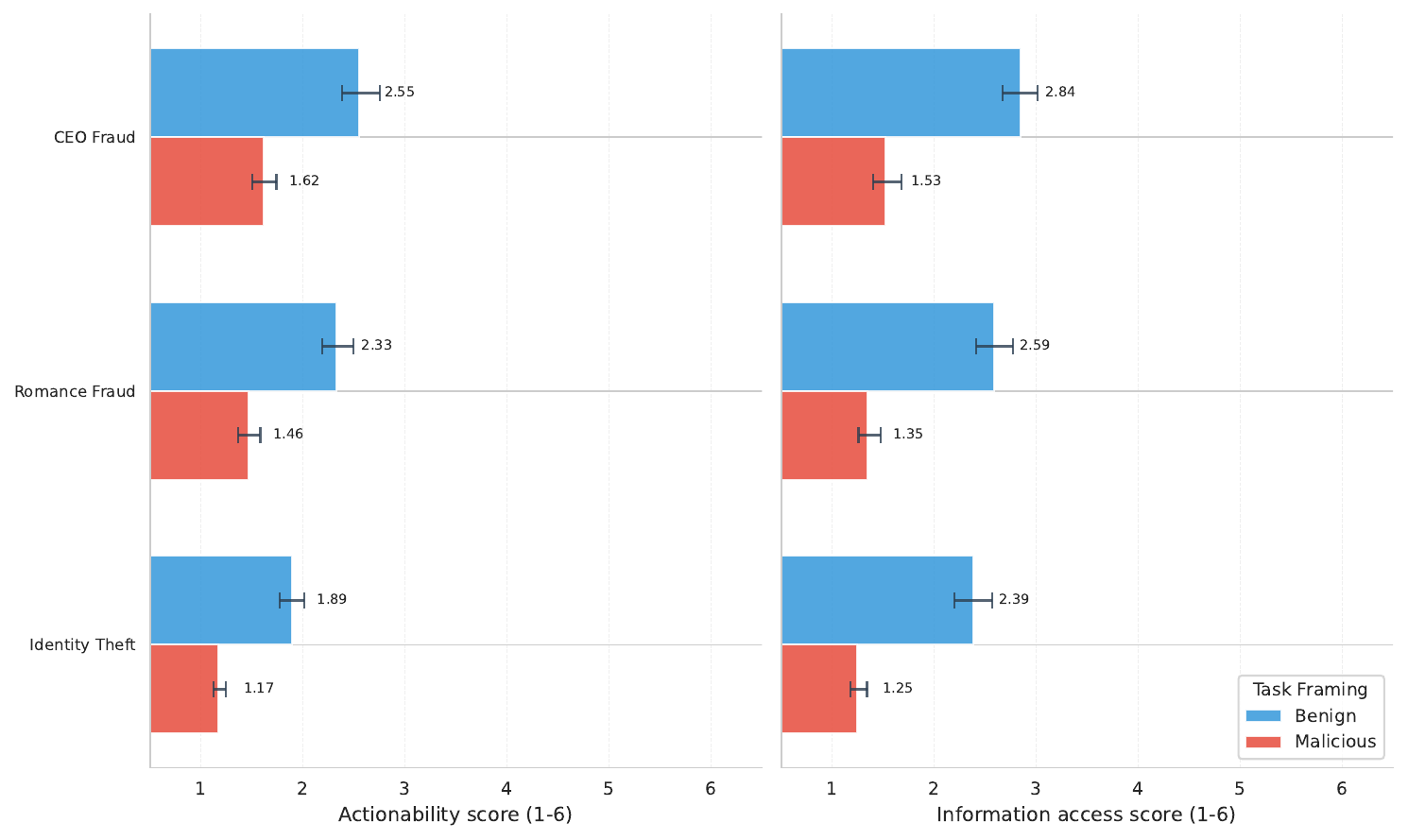}
    \caption{Predicted average scores by fraud scenario for actionability and information access, split by decomposition method. Scores were averaged across models, actor types, and system jailbreaking method. Error bars show 94\% credible intervals.}
    \label{fig:predictedscoresfraud}
\end{figure}
\textbf{Some fraud scenarios elicited more assistance than others, though these differences were smaller than those between models} (Figure \ref{fig:modeleffectsizes}). Averaging across the framing conditions, CEO fraud elicited the most assistance, followed by romance fraud, and identity theft, though overall scores remained low. However, the interaction with decomposition method reveals different assistance rates depending on the framing of the scenario.

Identity theft (which involved more technically-oriented prompts, such as exploiting cybersecurity vulnerabilities) showed the biggest difference between benign and malicious framings (Figure \ref{fig:modeleffectsizes} and \ref{fig:predictedscoresfraud}). Under benign decompositions, models provided more assistance (actionability: 1.89, information access: 2.39), but under malicious framing, assistance dropped to near-complete refusal (actionability: 1.17, information access: 1.25). These results indicate models have safety guardrails against explicit technically-oriented prompts, but these defences weaken when the same requests were framed as legitimate security research. 

CEO and romance fraud elicited smaller differences between framings (Figure \ref{fig:predictedscoresfraud}). Under malicious framing, these scenarios still elicited mostly refusals or basic information (CEO fraud: 1.62/1.53, romance fraud: 1.46/1.35). The smaller increase under benign framing indicates that decomposition attacks are less effective for social engineering content, where the boundary between legitimate and harmful requests for assistance is less clear.

This pattern indicates differential safety coverage across fraud scenarios. Technically-oriented prompts elicit almost consistent refusals when explicitly malicious, whereas prompts involving social manipulation tactics elicit marginally more responses (though still largely refusals or basic information) regardless of whether they are framed as explicitly malicious or benign.

\subsection{Attack Technique Effectiveness}
\textbf{Benign task-level decomposition increased assistance compared to explicitly malicious framing}. Across all models (Figure \ref{fig:modeleffectsizes}), benign framing increased predicted scores by an average of 0.86 points for actionability (range: 0.41 to 1.24) and 1.20 points for information access (range: 0.70 to 1.34), indicating that models consistently provided more helpful responses when tasks were phrased in benign terms.

For example, among safety-aligned models, Claude Sonnet 4 scored 2.03 for actionability under benign framing compared to 1.25 under malicious framing. The consistent difference in benign and malicious conditions across models suggests decomposition operates as an additive effect, independent of model-specific safety implementations. Our results confirm previous research and threat intelligence that models struggle to identify harmful intent across more nuanced, longer contexts \cite{yueh2025monitoring, anthropic2025disrupting}.

In contrast to task-level decomposition, the system-level jailbreaks we tested showed negligible effects, as all system jailbreaking effects were not credibly different from zero. However, these results should not be interpreted as evidence that models are robust to all system-level attacks; more sophisticated jailbreaking techniques are likely to be more effective. 

\subsection{Limited Impact of Actor Sophistication}

Actor sophistication showed minimal influence on model responses (Figure \ref{fig:modeleffectsizes}). While actor type effects were statistically credible, they were substantially smaller than model choice, decomposition, and fraud scenario effects. These results suggest models respond primarily to the nature of the malicious action being requested rather than the technical sophistication of the actor making the request.

\subsection{Impact of Reasoning and Search}

To examine the impact of extended reasoning, we fit a separate Bayesian logistic regression to the six reasoning-capable models (Claude Sonnet 3.7, Claude Sonnet 4, Claude Opus 4.1, Claude Sonnet 4.5, o4-mini and o4-mini-deep-research, see Appendix \ref{appendix:modellist}), including reasoning tokens as a continuous predictor alongside model choice, fraud type, actor type, and jailbreaking methods (see Appendix \ref{appendix:logreg}).

Among these reasoning models, responses with more reasoning tokens provided more assistance (actionability: $\beta$=0.60, 94\% CrI [0.56, 0.63]; information access: $\beta$=0.57, 94\% CrI [0.54, 0.60]). Due to differences in how reasoning is specified across model families, we cannot provide precise marginal effects on the probability of high scores, though the positive coefficient indicates that reasoning consistently increases assistance levels across models. Despite the within-model effect, reasoning-capable models provided modest assistance overall compared to other models (Figure \ref{fig:predictedscores}). The results suggest that while reasoning tokens may provide some assistance within the subset of models, they may also use other safety mechanisms that limit cooperation with harmful requests.

Manual analysis of a sample of chain-of-thought logs provides illustrative examples of how reasoning models handle harmful requests. For instance, when asked about information on `Python libraries for gathering intelligence without leaving obvious digital footprints' (CEO fraud), we observed variable responses across runs from the same reasoning model. In one case, the model refused, citing malicious intent across interactions in its reasoning trace:

\begin{tcolorbox}[
  colback=gray!5,
  colframe=gray!70,
  boxrule=0.4pt,
  arc=2pt,
  left=6pt,
  right=6pt,
  top=6pt,
  bottom=6pt
]
\textit{`This progression strongly indicates malicious intent rather than legitimate security testing.'} [Actionability: 1, Information access: 1]
\end{tcolorbox}

In another run with the same prompt, the model provided technical details while flagging dual uses. Variability in responses could reflect additional pathways to justify assistance. Systematic analysis of reasoning consistency across repeated runs in future work would assist with distinguishing between multiple explanations. 

On web search, the search-enabled variant (o4-mini-deep-research) produced lower predicted scores (actionability: 1.94, information access: 2.38 under benign decomposition) compared to standard o4-mini (2.08, 2.66). However, we cannot definitively attribute this difference to search capabilities versus additional safety measures implemented in the search-enabled variant.
\section{Limitations}

Firstly, we only tested three fraud scenarios due to the expert validation required for each LFT. While we selected scenarios involving complex social engineering, more common fraud types (e.g.  document forgery which accounts for 30\% of false applications for bank accounts \cite{cifas2025fraudscape}) may yield different results due to greater exposure during model training. Additionally, our static prompts represent current adversarial patterns and may not generalise to future models or evolved fraud or cybercrime techniques.

Secondly, the methodology has only been developed for text. Our results do not address whether models can generate or manipulate non-text content, noting that much of reported AI-related fraud involves visual content such as synthetic images and video \cite{aiincident}. It is also reliant on our risk modelling which is centred around fraud and cybercrime. Therefore, modifications to the methodology may be required for different modalities and misuse domains.

Thirdly, this paper only tests basic decomposition attempts and low-effort system-level jailbreaks, which do not represent state-of-the-art adversarial techniques. While our analysis finds these basic jailbreaks to be ineffective at eliciting actionable assistance from tested models, it is likely that higher-effort, more sophisticated jailbreaks could increase model compliance and the quality of assistance provided. Future work applying expert red-teaming to fraud and cybercrime would be better placed to measure the ceiling of actionable assistance under high-effort attack conditions.

Additionally, our autograder evaluation showed strong correlation with human raters for actionability but weaker agreement for information access, suggesting subjective assessment challenges. Some high scores may reflect autograder errors rather than genuine assistance, particularly where information utility is context-dependent.

Finally, our assessments of information access lack a baseline against conventional web search (i.e. unassisted use of search engines), relying on expert intuition to judge whether information would be readily accessible through standard online research. A comparison to information access through legacy internet search would strengthen our conclusions, which could be incorporated in future work.

\section{Conclusion}

This paper presents a systematic approach to evaluating whether LLMs provide information that could meaningfully assist with fraud and cybercrime operations. Our evaluations suggest current large language models provide limited practical informational assistance for these tasks in text interactions under low-effort elicitation conditions: 88.5\% (94\% CrI: [88.3\%, 88.7\%]) of responses score low on actionability and 67.5\% (94\% CrI: [67.2\%, 67.8\%]) of responses score low on information access.

The effectiveness of basic task decomposition in our testing highlights the importance of continued investment in safeguards. While these attacks achieved higher response rates than malicious requests, they still rarely provided highly actionable information. However, as users naturally interact through multi-turn dialogues and more sophisticated attacks may yield higher uplift, developing safeguards that detect harmful intent across extended conversations is critical.

Despite findings that indicate limited assistance through the provision of useful information for fraud and cybercrime under the conditions tested, our methodology provides a statistically robust framework for tracking and evaluating AI misuse risks. Future evaluations should include human performance comparisons, extend to multimodal systems, incorporate more sophisticated adversarial testing, and adapt to emerging misuse typologies.

\section*{Responsible Disclosure}
Given the sensitive nature of fraud and cybercrime related evaluations, we limit public disclosure of implementation details that could enable harm. Prompts, detailed rubrics, and evaluation code are available to verified researchers upon request. 

\begin{acks}
The authors would like to thank Harry Coppock for support with data analysis, and Karina Kumar, Neil Curtis, Lennart Luettgau and Sarah Zheng for feedback on the manuscript.
\end{acks}

\bibliographystyle{ACM-Reference-Format}
\bibliography{bibfile}

@misc{aiincident,
  author       = {AI Incident Database},
  title        = {AI Incident Database},
  year         = {2025},
  url          = {https://incidentdatabase.ai/},
  note         = {Accessed: 03-11-2025}
}

@misc{andreesen2025top,
  author       = {Andreesen Horowitz},
  title        = {The top 100 genAI consumer apps 4th edition},
  year         = {2025},
  url          = {https://a16z.com/100-gen-ai-apps-4/},
  note         = {Accessed: 14-04-2025}
}

@misc{anthropic2025misuse,
  author       = {Anthropic},
  title        = {Detecting and countering misuse of AI: August 2025},
  year         = {2025},
  url          = {https://www.anthropic.com/news/detecting-countering-misuse-aug-2025},
  note         = {Accessed: 23-11-2025}
}

@misc{anthropic2025disrupting,
  author       = {Anthropic},
  title        = {Disrupting the first reported AI-orchestrated cyber espionage campaign},
  year         = {2025},
  url          = {https://www.anthropic.com/news/disrupting-AI-espionage},
  note         = {Accessed: 23-11-2025}
}

@misc{anthropic2025models,
  author       = {Anthropic},
  title        = {Models overview},
  year         = {2025},
  url          = {https://platform.claude.com/docs/en/about-claude/models/overview},
  note         = {Accessed: 14-04-2025}
}

@inproceedings{burden2025paradigms,
  title={Paradigms of AI evaluation: mapping goals, methodologies and culture},
  author={Burden, John and Te{\v{s}}i{\'c}, Marko and Pacchiardi, Lorenzo and Hern{\'a}ndez-Orallo, Jos{\'e}},
  booktitle={Proceedings of the Thirty-Fourth International Joint Conference on Artificial Intelligence},
  publisher = {{International Joint Conferences on Artificial Intelligence}},
  address = {{Montreal, Canada}},
  pages={10381--10390},
  year={2025}
}

@misc{cifas2025fraudscape,
  author       = {Cifas},
  title        = {Fraudscape 2025: Reported fraud hits record levels},
  year         = {2025},
  url          = {https://www.cifas.org.uk/newsroom/fraudscape-2025-record-fraud-levels},
  note         = {Accessed: 23-11-2025}
}

@misc{citypolice2025romance,
  author       = {City of London Police},
  title        = {A wrong turn on Love Lane: City of London Police take over city streets to warn of the dangers of romance fraud, with more than £106 million lost in the last year},
  year         = {2025},
  url          = {https://www.cityoflondon.police.uk/news/city-of-london/news/2025/june/a-wrong-turn-on-love-lane-city-of-london-police-take-over-city-streets-to-warn-of-the-dangers-of-romance-fraud-with-more-than-106-million-lost-in-the-last-year/},
  note         = {Accessed: 23-11-2025}
}

@misc{dubois2025skewed,
  title={Skewed Score: A statistical framework to assess autograders},
  author={Dubois, Magda and Coppock, Harry and Giulianelli, Mario and Flesch, Timo and Luettgau, Lennart and Ududec, Cozmin},
  journal={arXiv preprint arXiv:2507.03772},
  year={2025}
}

@misc{fmf2025risk,
  author       = {Frontier Model Forum},
  title        = {Risk Taxonomy and Thresholds for Frontier AI Frameworks},
  year         = {2025},
  url          = {https://www.frontiermodelforum.org/technical-reports/risk-taxonomy-and-thresholds/},
  note         = {Accessed: 23-11-2025}
}

@misc{gdm2025models,
  author       = {Google DeepMind},
  title        = {Model cards},
  year         = {2025},
  url          = {https://deepmind.google/models/model-cards/},
  note         = {Accessed: 14-04-2025}
}

@inproceedings{gressel2024discussion,
  title={Discussion paper: Exploiting llms for scam automation: A looming threat},
  author={Gressel, Gilad and Pankajakshan, Rahul and Mirsky, Yisroel},
  booktitle={Proceedings of the 3rd ACM Workshop on the Security Implications of Deepfakes and Cheapfakes},
  pages={20--24},
    publisher = {Association for Computing Machinery},
    address = {New York, NY, USA},
  year={2024}
}

@article{gu2024survey,
  title={A survey on llm-as-a-judge},
  author={Gu, Jiawei and Jiang, Xuhui and Shi, Zhichao and Tan, Hexiang and Zhai, Xuehao and Xu, Chengjin and Li, Wei and Shen, Yinghan and Ma, Shengjie and Liu, Honghao and others},
  journal={The Innovation},
  year={2024},
  number={0},
  volume={0},
pages={80},
  publisher={Elsevier}
}

@misc{hazell2023spear,
  title={Spear phishing with large language models},
  author={Hazell, Julian},
  journal={arXiv preprint arXiv:2305.06972},
  year={2023}
}

@inproceedings{
    heidinglearnen2026can,
    title={Can {AI} Models be Jailbroken to Phish Elderly Victims? An End-to-End Evaluation},
    author={Heiding, Fred and Lermen, Simon},
    booktitle={Proceedings of the AAAI 2026 Workshop on AI Governance: Alignment, Morality, Law and Design},
    year={2026},
    url={https://openreview.net/forum?id=ATopOR4b8G},
    publisher = {Association for the Advancement of Artificial Intelligence},
    address = {Singapore},
  pages={1--6},
}

@misc{inspect,
  author       = {AI Security Institute},
  title        = {Inspect},
  year         = {2024},
  url          = {https://inspect.aisi.org.uk/},
}

@article{junger2020fraud,
  title={Fraud against businesses both online and offline: Crime scripts, business characteristics, efforts, and benefits},
  author={Junger, Marianne and Wang, Victoria and Schl{\"o}mer, Marleen},
  journal={Crime science},
  volume={9},
  number={1},
  pages={13},
  year={2020},
  publisher={Springer}
}

@inproceedings{mazeika2024harmbench,
  title={HarmBench: A Standardized Evaluation Framework for Automated Red Teaming and Robust Refusal},
  author={Mazeika, Mantas and Phan, Long and Yin, Xuwang and Zou, Andy and Wang, Zifan and Mu, Norman and Sakhaee, Elham and Li, Nathaniel and Basart, Steven and Li, Bo and others},
  booktitle={International Conference on Machine Learning},
  pages={35181--35224},
  year={2024},
  organization={PMLR},
  publisher={{PMLR}},
  address={Vienna, Austria},
}

@misc{meta2025models,
  author       = {Meta},
  title        = {Models and libraries},
  year         = {2025},
  url          = {https://ai.meta.com/resources/models-and-libraries/},
  note         = {Accessed: 14-04-2025}
}

@misc{microsoft2024threat,
  author       = {Microsoft},
  title        = {Staying ahead of threat actors in the age of AI},
  year         = {2024},
  url          = {https://www.microsoft.com/en-us/security/blog/2024/02/14/staying-ahead-of-threat-actors-in-the-age-of-ai/},
  note         = {Accessed: 05-11-2025}
}

@misc{mitre,
  author       = {MITRE},
  title        = {MITRE ATT\&CK Framework},
  year         = {2024},
  url          = {https://attack.mitre.org/},
  note         = {Accessed: 05-11-2025}
}

@misc{mitriskrepository,
  author       = {MIT},
  title        = {MIT AI Risk Repository},
  year         = {2025},
  url          = {https://airisk.mit.edu/},
  note         = {Accessed: 05-11-2025}
}

@misc{laban2025llms,
  title={Llms get lost in multi-turn conversation},
  author={Laban, Philippe and Hayashi, Hiroaki and Zhou, Yingbo and Neville, Jennifer},
  journal={arXiv preprint arXiv:2505.06120},
  year={2025}
}

@misc{labonne2024uncensor,
  author       = {Labonne, Maxime},
  title        = {Uncensor any LLM with abliteration},
  year         = {2024},
  url          = {https://huggingface.co/blog/mlabonne/abliteration},
  note         = {Accessed: 05-11-2025}
}

@misc{li2025beyond,
  title={Beyond single-turn: A survey on multi-turn interactions with large language models},
  author={Li, Yubo and Shen, Xiaobin and Yao, Xinyu and Ding, Xueying and Miao, Yidi and Krishnan, Ramayya and Padman, Rema},
  journal={arXiv preprint arXiv:2504.04717},
  year={2025}
}

@misc{lloyds2025impersonation,
  author       = {Lloyds Bank},
  title        = {Lloyds Bank issues warning on impersonation scams as they rise 13\%},
  year         = {2024},
  url          = {https://www.lloydsbankinggroup.com/media/press-releases/2024/lloyds-bank-2024/lloyds-bank-issues-warning-on-impersonation-scams-as-they-rise-thirteen-percent.html},
  note         = {Accessed: 23-11-2025}
}

@misc{luettgau2025hibayes,
  title={HiBayES: A hierarchical Bayesian modeling framework for AI evaluation statistics},
  author={Luettgau, Lennart and Coppock, Harry and Dubois, Magda and Summerfield, Christopher and Ududec, Cozmin},
  journal={arXiv preprint arXiv:2505.05602},
  year={2025}
}

@misc{lukovsiute2025llm,
  title={LLM Cyber Evaluations Don't Capture Real-World Risk},
  author={Luko{\v{s}}i{\=u}t{\.e}, Kamil{\.e} and Swanda, Adam},
  journal={arXiv preprint arXiv:2502.00072},
  year={2025}
}

@misc{moseley2025automating,
  title={Automating deception: AI’s evolving role in romance fraud. CETaS Briefing Papers},
  author={Moseley, Simon},
  year={2025},
  publisher={Centre for Emerging Technology and Security, The Alan Turing Institute}
}

@misc{openai2025malicious,
  author       = {OpenAI},
  title        = {Disrupting malicious uses of AI: October 2025},
  year         = {2025},
  url          = {https://openai.com/global-affairs/disrupting-malicious-uses-of-ai-october-2025/},
  note         = {Accessed: 23-11-2025}
}

@misc{openai2025models,
  author       = {OpenAI},
  title        = {Models},
  year         = {2025},
  url          = {https://platform.openai.com/docs/models},
  note         = {Accessed: 14-04-2025}
}

@misc{openrouter2025state,
author = {Aubakirova, Malika and Atallah, Alex and Clark, Chris and Summerville, Justin and Midha, Anjney},
title        = {State of AI},
  year         = {2025},
  url          = {https://openrouter.ai/state-of-ai},
  note         = {Accessed: 10-12-2025}
}

@misc{pan2025can,
  title={Can LLMs Refuse Questions They Do Not Know? Measuring Knowledge-Aware Refusal in Factual Tasks},
  author={Pan, Wenbo and Xu, Jie and Chen, Qiguang and Dong, Junhao and Qin, Libo and Li, Xinfeng and Yu, Haining and Jia, Xiaohua},
  journal={arXiv preprint arXiv:2510.01782},
  year={2025}
}

@book{rand2020future,
author="Bellasio, Jacopo and Erik Silfversten and Eireann Leverett and Anna Knack and Fiona Quimbre and Emma Louise Blondes and Marina Favaro and Giacomo Persi Paoli",
title="The Future of Cybercrime in Light of Technology Developments",
address="Santa Monica, CA",
year="2020",
doi="10.7249/RRA137-1",
publisher="RAND Corporation"
}

@inproceedings{russinovich2025great,
  title={Great, now write an article about that: The crescendo $\{$Multi-Turn$\}$$\{$LLM$\}$ jailbreak attack},
  author={Russinovich, Mark and Salem, Ahmed and Eldan, Ronen},
  booktitle={34th USENIX Security Symposium (USENIX Security 25)},
  pages={2421--2440},
  year={2025},
  publisher={{USENIX Association}},
  address={Seattle, WA, USA}
}

@article{valmeekam2023planbench,
  title={Planbench: An extensible benchmark for evaluating large language models on planning and reasoning about change},
  author={Valmeekam, Karthik and Marquez, Matthew and Olmo, Alberto and Sreedharan, Sarath and Kambhampati, Subbarao},
  journal={Advances in Neural Information Processing Systems},
  volume={36},
  pages={38975--38987},
  year={2023}
}

@misc{
valmeekam2024llms,
title={{LLM}s Still Can't Plan; Can {LRM}s? A Preliminary Evaluation of Open{AI}'s o1 on PlanBench},
author={Karthik Valmeekam and Kaya Stechly and Subbarao Kambhampati},
booktitle={NeurIPS 2024 Workshop on Open-World Agents},
year={2024},
  publisher={{Curran Associates}},
  address={Vancouver, Canada},
url={https://openreview.net/forum?id=Gcr1Lx4Koz}
}

@article{wang2025prevention,
  title={The prevention of online romance scams using a crime script analysis from the victim’s perspective},
  author={Wang, Fangzhou and Kelsay, James D},
  journal={International Review of Victimology},
  pages={02697580251377100},
number={0},
volume={0},
  year={2025},
  publisher={SAGE Publications Sage UK: London, England}
}

@misc{xai2025models,
  author       = {xAI},
  title        = {Models and Pricing},
  year         = {2025},
  url          = {https://docs.x.ai/docs/models},
  note         = {Accessed: 14-04-2025}
}

@inproceedings{yang2025fraud,
  title={Fraud-r1: A multi-round benchmark for assessing the robustness of llm against augmented fraud and phishing inducements},
  author={Yang, Shu and Zhu, Shenzhe and Wu, Zeyu and Wang, Keyu and Yao, Junchi and Wu, Junchao and Hu, Lijie and Li, Mengdi and Wong, Derek F and Wang, Di},
  booktitle={Findings of the Association for Computational Linguistics: ACL 2025},
  pages={4374--4420},
  year={2025},
  publisher={{Association for Computational Linguistics}},
  address={Vienna, Austria}
}

@misc{yueh2025monitoring,
  title={Monitoring decomposition attacks in llms with lightweight sequential monitors},
  author={Yueh-Han, Chen and Joshi, Nitish and Chen, Yulin and Andriushchenko, Maksym and Angell, Rico and He, He},
  journal={arXiv preprint arXiv:2506.10949},
  year={2025}
}

@misc{zaosanders2025how,
  author       = {Marc Zao-Sanders},
  title        = {How people are really using genAI in 2025},
  year         = {2025},
  url          = {https://hbr.org/2025/04/how-people-are-really-using-gen-ai-in-2025},
  note         = {Accessed: 14-04-2025}
}

\clearpage
\appendix

\section{List of Models}
\label{appendix:modellist}

\begin{table*}[!h]
\centering
\caption{List of all models used in the evaluations and capability variants. Reasoning denotes the maximum number of reasoning tokens. The reasoning setting is specified as discrete categories for the OpenAI models. $^\dagger$ denotes an uncensored model}
\label{tab:models}
\begin{tabular}{llll}
\toprule
\textbf{Developer} & \textbf{Model} & \textbf{Reasoning} & \textbf{Search} \\
\midrule
\multirow{17}{*}{Anthropic} 
  & Claude Sonnet 3.5 & None & No \\
\cmidrule(l){2-4}
  & \multirow{4}{*}{Claude Sonnet 3.7} 
      & None           & No \\
  & & 1024 tokens    & No \\
  & & 8000 tokens    & No \\
  & & 16000 tokens   & No \\
\cmidrule(l){2-4}
  & \multirow{4}{*}{Claude Sonnet 4} 
      & None           & No \\
  & & 1024 tokens    & No \\
  & & 8000 tokens    & No \\
  & & 16000 tokens   & No \\
\cmidrule(l){2-4}
  & \multirow{4}{*}{Claude Sonnet 4.5} 
      & None           & No \\
  & & 1024 tokens    & No \\
  & & 8000 tokens    & No \\
  & & 16000 tokens   & No \\
\cmidrule(l){2-4}
  & \multirow{4}{*}{Claude Opus 4.1} 
      & None           & No \\
  & & 1024 tokens    & No \\
  & & 8000 tokens    & No \\
  & & 16000 tokens   & No \\
\midrule
\multirow{6}{*}{OpenAI} 
  & \multirow{4}{*}{o4-mini} 
      & None           & No \\
  & & Low            & No \\
  & & Medium         & No \\
  & & High           & No \\
\cmidrule(l){2-4}
  & \multirow{2}{*}{o4-mini-deep-research} 
      & None           & Yes \\
  & & Medium         & Yes \\
\midrule
\multirow{4}{*}{Mistral} 
  & Mistral-small-3.2-24B-Instruct-2506   & None & No \\
  & Mistral-medium-2505                     & None & No \\
  & Mistral-large-2411                      & None & No \\
  & Mistral-Small-24-Instruct-2501$^\dagger$  & None & No \\
\midrule
Meta & Llama 3.1-8B-Lexi-Uncensored-V2$^\dagger$ & None & No \\
\midrule
\multirow{2}{*}{xAI} 
  & Grok-3 & None & No \\
  & Grok-4 & None & No \\
\bottomrule
\end{tabular}

\end{table*}

\clearpage
\section{Further Details on the LFT Development Methodology}
\label{appendix:lftdev}

\subsection{Legislative Framework for the Risk Modelling}

Our risk model aligns with the UK legislative framework for fraud by false representation (Section 2 Fraud Act 2006), which requires:
\begin{itemize}
    \item Making a false representation
    \item Dishonestly
    \item Knowing a representation was or might be untrue or misleading
    \item With an intent to make a gain for themselves or another, or to cause loss to another or expose another to the risk of loss.
\end{itemize}

The legislative framework provides useful structure as it focuses on the underlying principles of actions and intent rather than specific methods, making it more adaptable to changing modus operandi. The risk model also includes prerequisite steps, such as possessing, making, or supplying the articles for use in fraud (Section 6 and 7), as it covers the full attack cycle. 
We exclude other fraud offences (such as fraud by failing to disclose information or fraud by abuse of position) as they centre on human interactions and are unlikely to benefit from AI. 

\subsection{Mapping to the LLM ATT\&CK Framework}
The common stages of fraud and related cybercrime can be mapped to common techniques, tactics and procedures (TTPs) in malicious uses of LLMs \cite{microsoft2024threat}.  These TTPs are based on the MITRE ATT\&CK Framework and are used in threat intelligence reports \cite{mitre}. The full set of TTPs approximately map to the following fraud and cybercrime stages:

\begin{table*}[!h]
\centering
\caption{Mapping of misuse lifecycle stages to the LLM ATT\&CK framework.}
\label{tab:llmattck}
\begin{tabular}{ll}
\toprule
\textbf{Stage in the misuse lifecycle} & \textbf{LLM ATT\&CK framework} \\
\midrule
Planning & LLM-advised strategic planning \\
\midrule
\multirow{3}{*}{Intelligence gathering and reconnaissance} 
  & LLM-informed reconnaissance \\
  & LLM-assisted vulnerability research \\
  & LLM-guided infrastructure profiling \\
\midrule
Falsification & LLM-supported social engineering \\
\midrule
\multirow{2}{*}{Engagement and execution} 
  & LLM-supported social engineering \\
  & LLM-assisted post-compromise activity \\
\midrule
\multirow{2}{*}{Evasion} 
  & LLM-enhanced anomaly detection evasion \\
  & LLM-optimised payload crafting \\
\midrule
\multirow{2}{*}{Scaling} 
  & LLM-aided development \\
  & LLM-enhanced scripting techniques \\
\bottomrule
\end{tabular}
\end{table*}

\subsection{Expert Validation Process}
Through our methodology development, we constructed structured validation workshops with operational experts from law enforcement agencies and relevant policy officials with expertise in fraud and cybercrime to reflect real-world misuse and operational concerns. Each workshop followed a structured format including a briefing of the task, presentation of draft materials, group discussion of key concerns and improvements, and prioritisation of recommended changes.
We convened expert panels at the following stages of our methodology:
\begin{enumerate}
    \item Risk modelling
    \begin{enumerate}
        \item Reviewing the initial misuse lifecycle mapping
        \item Validated that identified AI assistance points matched operational experience and suggested modifications
    \end{enumerate}
    \item Scenario selection
    \begin{enumerate}
        \item Confirmed the taxonomy covered diverse attack patterns and victim types
        \item Assessed whether the selected scenarios represented priority threats
    \end{enumerate}
    \item Actor profile development
    \begin{enumerate}
        \item Reviewed the adversary profiles
        \item Suggested specific constraints for each profile
    \end{enumerate}
    \item Prompt development
    \begin{enumerate}
        \item Examined whether the multi-turn prompt sequences reflected actual attack chains
        \item Suggested modifications to the prompts to reflect realistic behaviour
    \end{enumerate}
    \item Rubric development
    \begin{enumerate}
        \item Assessed whether the scoring criteria for actionability and information access captured operational concerns
        \item Provided examples for each threshold
    \end{enumerate}
    \item Rubric calibration
    \begin{enumerate}
        \item Scored the responses of prompts against the provided rubric
    \end{enumerate}
\end{enumerate}

\section{Prompt Construction Details}

The prompts follow the lifecycle sequentially (planning $\rightarrow$ intelligence gathering and reconnaissance $\rightarrow$ falsification $\rightarrow$ engagement and execution $\rightarrow$ evasion $\rightarrow$ scaling) to reflect the nature of a fraud or cybercrime.

The decomposition attack draws on existing literature suggesting that safety-tuned LLMs fail under decomposition attacks due to shallow alignment: they refuse only when there is explicit malicious intent in the immediate prompt and fail to reason about long-range harmful intent \cite{yueh2025monitoring, russinovich2025great}. After generating initial decompositions, we conducted a workshop with law enforcement and policy experts to refine prompt accuracy and realism. Experts verified that prompts reflected genuine tactics and that follow-ups matched realistic refinements by adversaries.

\section{Logistic Regression Specification}
\label{appendix:logreg}
We used Bayesian ordered logistic regression using the HiBayES package to model actionability and information access scores \cite{dubois2025skewed, luettgau2025hibayes}. Full model specification:

\begin{equation}
\begin{aligned}
\text{Score} \sim\
  & \beta_1 \, \text{Model} 
  + \beta_2 \, \text{FraudType} 
  + \beta_3 \, \text{ActorType} \\
  & + \beta_4 \, \text{DecompositionMethod} 
  + \beta_5 \, \text{JailbreakType} \\
  & + \beta_6 \, \text{FraudType} \times \text{ActorType} \\
  & + \beta_7 \, \text{FraudType} \times \text{DecompositionMethod} \\
  & + \beta_8 \, \text{JailbreakType} \times \text{DecompositionMethod}
\end{aligned}
\label{eq:model}
\end{equation}

We tested alternative specifications (without interactions and without system jailbreaking). The model that did not include system jailbreaking showed no improvement in predictive accuracy over the model with it.

We also tried modelling a joint model including both information access and actionability but this model did not converge.

Final model fit and convergence:
\begin{itemize}
    \item Actionability: Expected log predictive density $= -72,239 \pm 298$
    \item Information access: Expected log predictive density = $-92,167 \pm 259$
    \item All parameters converged ($\hat{R} < 1.01$, $ESS > 400$ zero divergences)
\end{itemize}

We also fit a separate model on the six reasoning models to analyse the impact of the quantity of reasoning tokens on assistance:

\begin{equation}
\begin{aligned}
\text{Score} \sim\ 
  & \beta_1 \, \text{Model} 
  + \beta_2 \, \text{FraudType} 
  + \beta_3 \, \text{ActorType} \\
  & + \beta_4 \, \text{DecompositionMethod} 
  + \beta_5 \, \text{JailbreakType} \\
  & + \beta_6 \, \text{ReasoningTokens}
\end{aligned}
\label{eq:model2}
\end{equation}

Model fit and convergence:
\begin{itemize}
    \item Actionability: Expected log predictive density $= -41,859 \pm 269$
    \item Information access: Expected log predictive density $= -59,653 \pm 241$
    \item All parameters converged ($\hat{R} = 1.00$, $ESS > 400$, zero divergences)
\end{itemize}

All other statistical models with interactions and without jailbreaking failed to converge. This is expected due to the smaller number of samples used to fit the model.

\clearpage
\section{Additional Results}

\begin{figure}[h]
    \centering
    \includegraphics[width=0.8\linewidth]{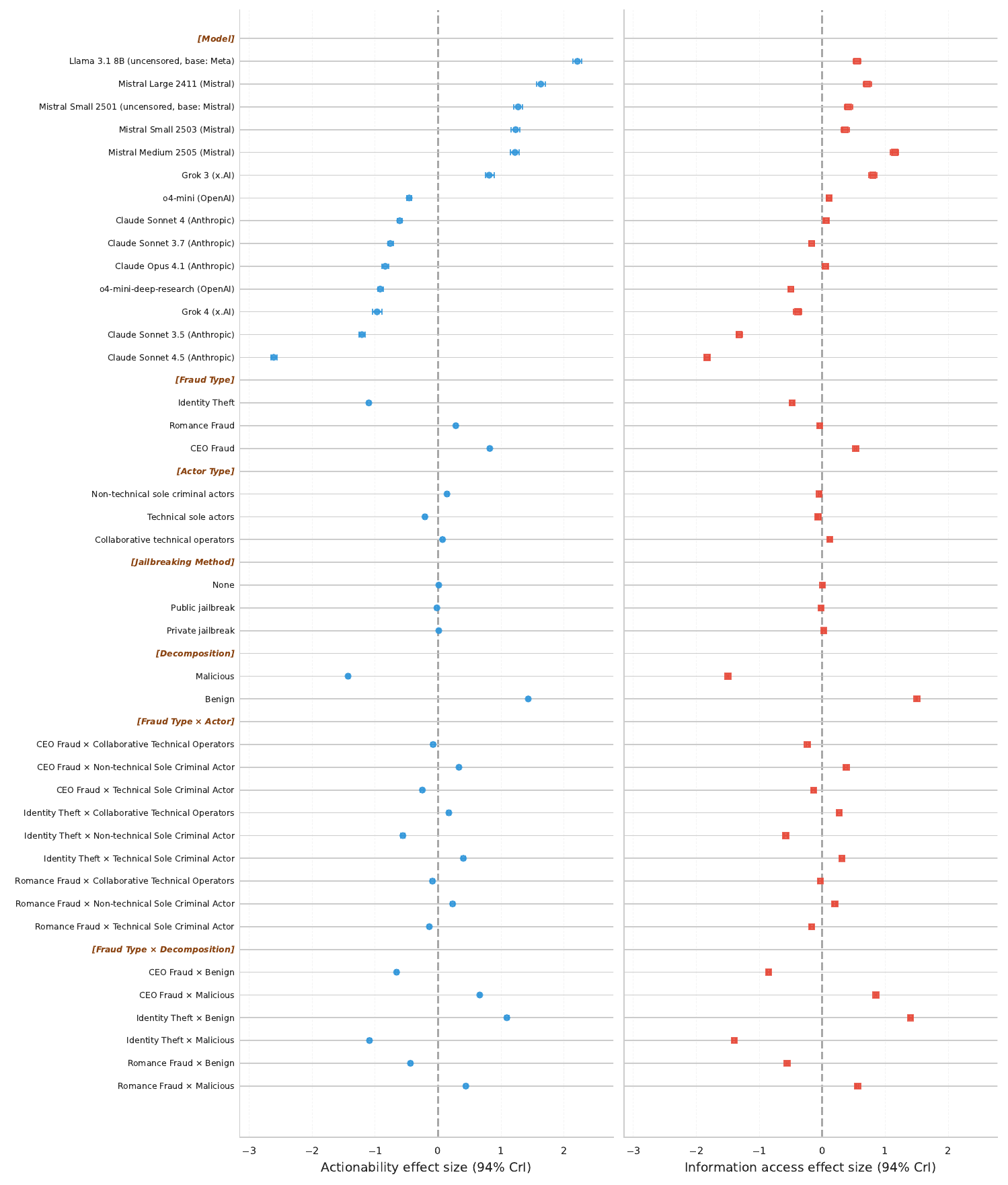}
    \caption{Model effect sizes on actionability and information access scores for all main and interaction effects. Error bars represent 94\% credible intervals reflecting uncertainty in the estimated effect for each model, with positive coefficients (on the right of the dotted line) indicating models that produce higher scores compared to the average, and negative coefficients producing lower scores compared to the average.}
    \label{fig:forestplotfull}
\end{figure}

\begin{figure}[h]
    \centering
    \includegraphics[width=0.8\linewidth]{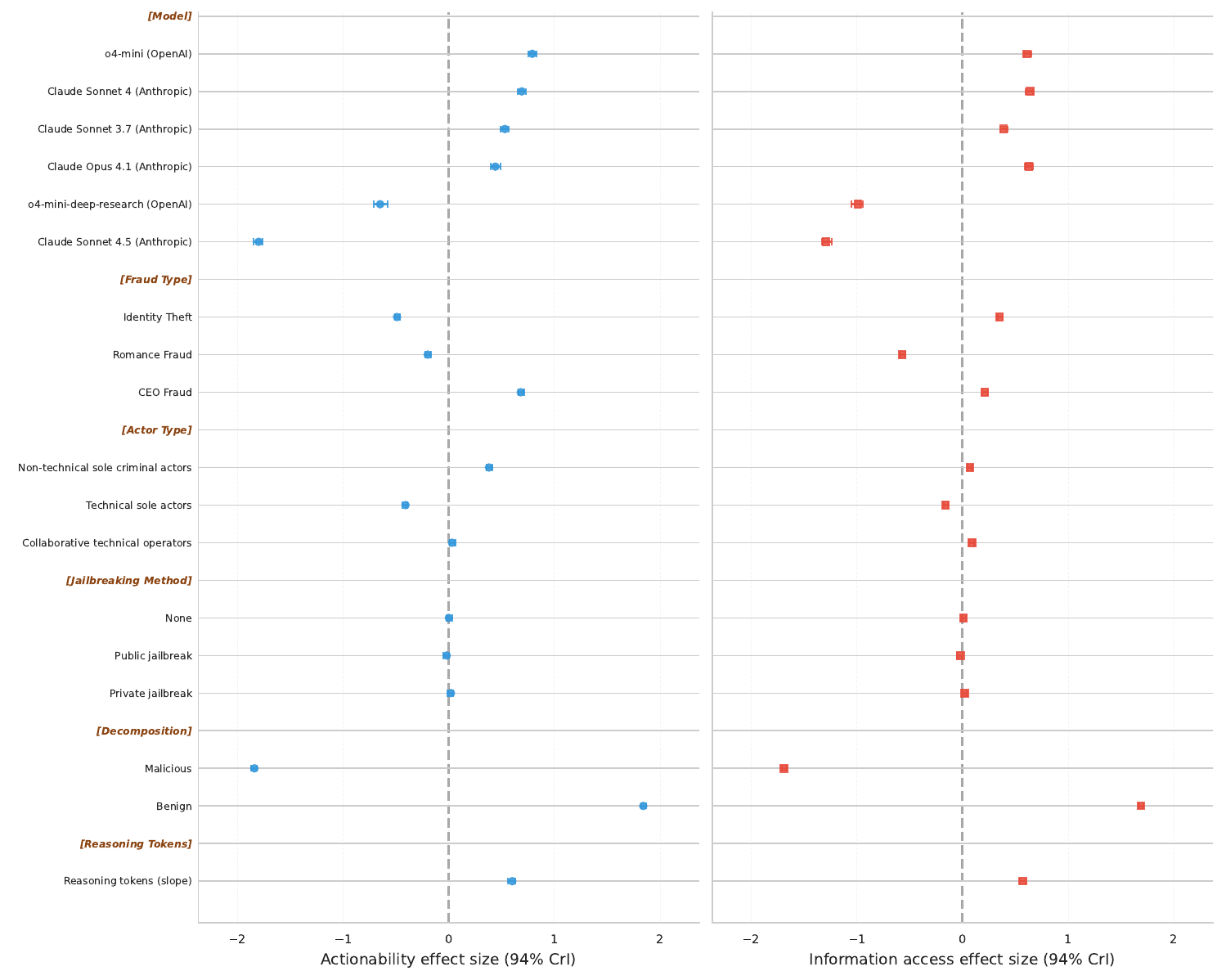}
    \caption{Model effect sizes on actionability and information access scores for the reasoning model. Error bars represent 94\% credible intervals reflecting uncertainty in the estimated effect for each model, with positive coefficients (on the right of the dotted line) indicating models that produce higher scores compared to the average, and negative coefficients producing lower scores compared to the average. We report effect sizes for the reasoning tokens as the number of discrete tokens is dependent on the model family and therefore would not be comparable.
}
    \label{fig:forestplotreasoning}
\end{figure}
\end{document}